\documentclass[10pt,twocolumn,letterpaper]{article}

\usepackage[pagenumbers]{iccv}

\usepackage{amsmath,amsfonts}
\usepackage{array}
\usepackage{textcomp}
\usepackage{stfloats}
\usepackage{url}
\usepackage{verbatim}
\usepackage{graphicx}
\def\BibTeX{{\rm B\kern-.05em{\sc i\kern-.025em b}\kern-.08em
    T\kern-.1667em\lower.7ex\hbox{E}\kern-.125emX}}
\usepackage{balance}

\usepackage{epsfig}

\usepackage{amssymb}
\usepackage{amsthm}

\usepackage{graphbox} 

\usepackage[dvipsnames]{xcolor}

\usepackage{tabstackengine}
\setstackEOL{\cr}
\usepackage{booktabs}
\usepackage[utf8]{inputenc}
\usepackage{lscape}
\usepackage{subcaption}
\usepackage{longtable}

\usepackage{makecell}
\usepackage{algorithm}
\usepackage{algorithmic}
\usepackage{multirow}

\usepackage{rotating}

\usepackage{parnotes}

\usepackage{enumitem}

\usepackage[numbers]{natbib}

\definecolor{iccvblue}{rgb}{0.21,0.49,0.74}
\usepackage[pagebackref,breaklinks,colorlinks,allcolors=iccvblue]{hyperref}

\begin{document}
\title{RSR-NF: Neural Field Regularization by Static Restoration Priors\\for Dynamic Imaging}

\author{Berk Iskender$^{1,*}$ $\quad$ Sushan Nakarmi$^2$ $\quad$ Nitin Daphalapurkar$^2$ $\quad$ Marc L. Klasky$^2$ $\quad$ Yoram Bresler$^1$ \\
$^1${\small University of Illinois at Urbana-Champaign, IL, US} $\quad$ {\small $^2$Los Alamos National Laboratory, NM, US} 
}

\maketitle
\makeatletter{\renewcommand*\@makefnmark{}
\footnotetext{$^*$\textit{Corresponding author: {\tt\scriptsize berk.iskender@analog.com}. Currently at Analog Garage, Analog Devices, Inc., Boston, MA, US.}}\makeatother}

\makeatletter{
\renewcommand*\@makefnmark{}
\footnotetext{
This research was supported in part by Los Alamos National Labs under Subcontract No. 599416/CW13995.
}\makeatother}

\setlength{\abovedisplayskip}{8pt}
\setlength{\belowdisplayskip}{8pt}

\begin{abstract}

Dynamic imaging involves the reconstruction of a spatio-temporal object at all times using its undersampled measurements. In particular, in dynamic computed tomography (dCT), only a single projection at one view angle is available at a time, making the inverse problem very challenging. Moreover, ground-truth dynamic data is usually either unavailable or too scarce to be used for supervised learning techniques. To tackle this problem, we propose RSR-NF, which uses a neural field (NF) to represent the dynamic object and, using the Regularization-by-Denoising (RED) framework, incorporates an additional \textit{static} deep spatial prior into a variational formulation via a learned restoration operator. We use an ADMM-based algorithm with variable splitting to efficiently optimize the variational objective.
We compare RSR-NF to three alternatives: NF with only temporal regularization; a recent method combining a partially-separable low-rank representation with RED using a denoiser pretrained on static data; and a deep-image prior-based model. The first comparison demonstrates the reconstruction improvements achieved by combining the NF representation with static restoration priors, whereas the other two demonstrate the improvement over state-of-the art techniques for dCT.
\end{abstract}
\section{Introduction}
\subsection{Dynamic Imaging}
Dynamic imaging addresses the ill-posed problem of reconstructing a time-varying object from its undersampled measurements. In the extreme case, these measurements are time-sequential, meaning that only a single measurement is available at any given time instant. {This problem arises in many  areas of science and engineering, including} dynamic computed tomography (dCT), \cite{willis95, mohan2015timbir, zang2018space, huangUnetbasedDeformationVector2020, madestaSelfcontainedDeepLearningbased2020, majee2021multi, iskender2022dynamic, iskender2022dynamic_long, iskender2023factorized, Iskender_2023_ICCV, iskender2023red, iskender2023dynamic, iskender2023redtci}, dynamic MRI (dMRI) \cite{liang2007spatiotemporal, Haldar2010, zhao2012image, babu2023fast, yoo2021time, zou2021dynamic, ahmed2022dynamic}, photoacoustic CT (PACT) \cite{lozenski2022memory, lozenski2024proxnf}, and dynamic 3D scene representation \cite{xian2021space, li2021neural, du2021neural, park2021nerfies, park2021hypernerf, Gao-ICCV-DynNeRF, wang2021neural}. 
 
Dynamic imaging is more challenging compared to its static version since different measurements in the undersampled set belong to essentially different objects. Thus, the traditional algorithms for static reconstruction lead to severe artifacts, which is particularly evident for time-sequential acquisition.

In this paper, focusing on the dCT problem, we introduce a new approach and algorithm, capable of reliable reconstruction even in the extreme case of fully time-sequential acquisition where a single projection at a single view angle at a time is acquired from a highly dynamic object. The experiments in this paper address dCT, but the proposed approach is applicable to other dynamic imaging modalities. To limit the scope of the discussion though, we restrict here our review of previous work mostly to dCT and to the closely related problem of dMRI.

\subsection{Related Previous Work}
\label{sec:intro_prev_work}
To address the challenging ill-posed problem in dCT and dMRI, {a common general approach has been to represent or model the underlying dynamic object using a reduced number of degrees of freedom.}
See \cite{iskender2023red, iskender2023redtci} for a detailed overview of previously proposed techniques.
{We highlight below those that are most relevant to the work in this paper.} 

\textbf{Low rank methods.}
{These methods} model the underlying object as low-rank in one of two ways: (i) either enforcing this as a hard constraint {via a partially-separable model (PSM), which uses a low-rank factorization of the object into spatial and temporal factors} \cite{liang2007spatiotemporal, pedersen2009k, Haldar2010, haldar2011low, zhao2011further, zhao2012image, lingala2013blind, iskender2022dynamic, iskender2022dynamic_long, djebra2022manifold, iskender2023factorized}, recently augmented by pre-learned spatial priors \cite{Iskender_2023_ICCV, iskender2023red, iskender2023dynamic, iskender2023redtci}; or (ii) promoting the low rank structure by including {nuclear or more general Schatten-$p$ norm} penalties in the optimization objective, often with sparsity regularization in a transform domain \cite{lingala2011accelerated, majumdar2015learning, Trzasko2013}. 
Other such methods decompose the object into the sum of low-rank and sparse components, again promoting low-rankness via nuclear or Schatten-$p$ norms \cite{chiew2015k, otazo2015low}, with a recent method also decomposing the mean signal as a third component \cite{babu2023fast}.

Recent low-rank methods with pre-learned spatial priors \cite{Iskender_2023_ICCV, iskender2023red, iskender2023dynamic, iskender2023redtci} were shown to provide considerable improvements over  other low-rank alternatives with simpler priors, and over
Deep Image Prior (DIP)-based techniques \cite{yoo2021time}. However, {as demonstrated by direct evaluation in Section~\ref{subsec:results},} they may not provide sufficiently accurate reconstructions when the underlying dynamic object is high-rank. 

\textbf{Motion estimation and compensation.}
Another group of algorithms tries to recover the object using motion estimation and compensation methods \cite{lingala2014deformation, chen2014motion, yoon_motion_2014, tolouee2018nonrigid}. Estimating a continuous motion field, these methods assume fixed total density and they are unable to represent topological changes {(See Section~\ref{subsec:proposed} for further discussion of these limitations.)}

\textbf{Deep image prior.}
Recent dynamic imaging algorithms apply the deep learning (DL)-based deep image prior (DIP) \cite{ulyanov2018deep} idea to spatio-temporal signals to represent the dynamic object at each time instant using a low-dimensional latent representation\cite{yoo2021time, zou2021dynamic}. The latent representations {drive} a generator neural network (NN) {producing} the estimated object.  Here the NN parameters and potentially the latent representations \cite{hyder2019generative} are estimated in an unsupervised manner. Later work combines a low-rank PSM representation with the DIP framework \cite{ahmed2022dynamic}. {Drawbacks of the DIP methods include the need for early stopping  to avoid overfitting and degradation which is difficult without a ground-truth oracle, difficult convergence requiring multiple random initializations, and lack of a mechanism to incorporate a prior other than indirectly and coarsely through the choice of deep NN architecture.}

\textbf{Neural fields.}
Neural fields (NFs), 
or implicit neural representations, 
have been proposed lately as DL-based continuous representations of {natural scenes and static and dynamic objects in a wide range of applications \cite{xie2022neural}.}
{In particular, applications in computed imaging include}
static CT \cite{sun2021coil, zang2021intratomo, shen2022nerp, song2023piner} and MRI \cite{shen2022nerp}, dCT \cite{reed2021dynamic, zhang2023dynamic, birklein2023neural} and dMRI \cite{feng2022spatiotemporal, huang2023neural, spieker2023iconik, kunz2023implicit}, dynamic PACT \cite{lozenski2022memory, lozenski2024proxnf}, cryo-EM \cite{zhong2021cryodrgn}, and medical imaging in general \cite{wang2024neural}. 
{A NF representation is particularly effective in dynamic imaging problems since it is essentially resolution-free, and the number of parameters to represent, and also the memory required to store the object do not scale with the spatio-temporal resolution.}


For dCT, 
{several methods   
\cite{reed2021dynamic,zhang2023dynamic,shao2023dynamic,birklein2023neural} decouple the reconstruction into recovering a reference (``template") cross-section or volume via a spatial NF, and estimating a motion field to represent the dynamics by warping the template.}
{These methods are subject to 
 limitations similar to those of other motion field-based techniques (see above).} 
{For dynamic PACT, some {NF}-based methods} \cite{lozenski2022memory,lozenski2024proxnf}  use a partition of unity NN (POUnet) \cite{lee2021partition} as the NF architecture to represent the dynamic object, %
{which is claimed to improve performance over that by a standard NF. In our experiments in Section~\ref{subsec:results}, we compared the representation power of a POUnet with that of a standard NF with positional encoding, and did not encounter a notable difference.}

{In the {computed dynamic} imaging applications discussed above,}
similar to PSM or DIP-based methods, most of the NF-based models are {fitted} in an unsupervised manner, using only the measurement data to estimate the NN parameters. Also, regularization in addition to the NF, if any, is in the form of traditional, fixed regularizers. This misses the opportunity to use more powerful learning-based targeted regularization.

{
{Sharing some elements with the proposed approach,} a recent method for \textit{static} sparse-view CT \cite{vo2024neural} combines a NF representation for the object with a learned regularizer using the RED framework \cite{vo2024neural}.}
{Due to high computational load}, the regularization is performed on a subset of blocks of the estimated object for each update of the NF parameters. {However, even with this strategy, sweeping blockwise over the entire object remains computationally heavy.}
The reconstruction {quality is similar to that of the comparison benchmarks.}
{Furthermore,} despite using in RED an artifact removal NN as in RARE \cite{liu2020rare} instead of a denoiser, the NN is trained in a supervised manner.
Unfortunately, such an approach is not feasible for the majority of dynamic imaging problems, including dMRI and dCT where there is no ground-truth dynamic data.

\subsection{Proposed Approach}
\label{subsec:proposed}
To overcome the {limitations} of previous methods, while still utilizing learned spatial priors as in \cite{Iskender_2023_ICCV, iskender2023red, iskender2023dynamic, iskender2023redtci}, we propose the reconstruction algorithm {\textit{R}egularization by \textit{S}tatic \textit{R}estoration priors} with a {\textit{N}eural \textit{F}ield  (RSR-NF) for dynamic imaging}, which uses {a neural field to provide a highly expressive yet parsimonious} representation for the spatio-temporal object, and combines it with a {smoothness prior for temporal regularization,} and with a pre-learned spatial {restoration} prior {using the RED framework} \cite{romano2017little}, {to enable recovery in critically undersampled scenarios.}  
{The latter feature} allows RSR-NF to exploit the performance of a {restoration network pre-trained on available \textit{static} ground-truth data}.
To solve the resulting non-convex optimization problem, we use an ADMM algorithm with an efficient fixed-point update for the spatial regularizer. 

{RSR-NF offers several advantages over previous NF-based reconstruction methods {for dynamic imaging}. Importantly, unlike \cite{zhang2023dynamic},} RSR-NF does not require ground-truth dynamic data. Furthermore, in contrast to the NF-based algorithms such as \cite{reed2021dynamic, zhang2023dynamic, birklein2023neural, shao2023dynamic} 
that recover a static template and the corresponding motion field to warp it, RSR-NF does not {make the assumption of} a fixed total density, which may be restrictive (e.g, injecting a contrast agent to the field-of-view, or cross-slice motion while imaging a fixed slice can violate this assumption). Another issue with the motion field-based methods is their inability to represent topological changes, which require discontinuity in the representation \cite{park2021hypernerf}. 
Finally, {compared to the NF-based approach in \cite{lozenski2022memory}, which uses simpler regularizers,} the variational objective in RSR-NF is regularized with more accurate, pre-learned spatial priors.

\textbf{Contributions.} 
\textbf{(1)} To the best of our knowledge, RSR-NF is the first method  {for dynamic object reconstruction}, to combine NF-based object representation with a pre-learned, {static spatial prior.}
\textbf{(2)} RSR-NF does not require \textit{spatio-temporal} training data, which is typically unavailable or very difficult to obtain. {RSR-NF is therefore applicable to a wide range of dynamic inverse problems.}
\textbf{(3)} Unlike {methods such as \cite{liu2022recovery}, which use a deep NN regularizer, the}
optimization scheme in RSR-NF avoids costly {backpropagation computations} through the deep restoration NN {for updating the NF parameters.}

\section{{The dCT Imaging Inverse Problem}
\label{sec:dct_prob}}
In dCT the objective is to reconstruct a spatio-temporal object $f(\boldsymbol{x}, t)$, $\boldsymbol{x} \in \mathbb{R}^{d}$ {(with $d=2$ or $d=3$ for individual slice or volumetric reconstruction)}, from its projections
\begin{align}
    \label{eq:dyn_tomo_prob}
    g(\cdot, \theta, t) = \mathcal{R}_\theta \{ f(\boldsymbol{x}, t) \} + \eta(\cdot, \theta, t)
\end{align}
obtained using the Radon transform operator $\mathcal{R}_\theta$ at view angle $\theta$ and corrupted by additive noise $\eta$. 
Considering the extreme case of time-sequential sampling with only one projection acquired at each time instant with uniform temporal spacing, the set of measurements is
\begin{align}
    \label{eq:acquired_meas}
    \{ g(s, \theta_p, t_p) \}_{p=0}^{P-1}, \quad \forall s, t_p = p \Delta_t,
\end{align}
where $s$ is the offset of the line of integration from the origin (i.e., detector position), and $P$ is the total number of projections (and temporal samples) acquired. Sufficient sampling of the variable $s$ is assumed, and it is suppressed in the following notation. 
The angular sampling scheme, $\{ \theta_p \}_{p=0}^{P-1}$, with $\theta_p \in [0, 2\pi]$, is a free design parameter.

This ill-posed inverse problem is particularly challenging since, in the time-sequential setting, each projection belongs to a different object. 
This is a significant drop from the required more than $C$ projections for the artifact-free reconstruction of a static object of a diameter of $C$-pixels. 
To address this issue, several techniques \cite{zang2018space, mohan2015timbir, majee2021multi} cluster temporally adjacent projections and {model} the object as static during their acquisition. However, this reduces the temporal resolution, and any departure from this assumption (as in the case of time-sequential sampling) {gives rise} to modeling inaccuracies in the measurement fidelity term, leading to reconstruction errors.

{On the other hand, in certain cases, pre- and post-motion static reconstructions of the dynamic object, or reconstructions of similar static objects are available and can be exploited to improve the estimated solution accuracy. {This is the scenario we address.}
\section{Proposed Method: RSR-NF {(Figure~\ref{fig:framework})} }

\begin{figure}
    \centering
    \includegraphics[width=\linewidth]{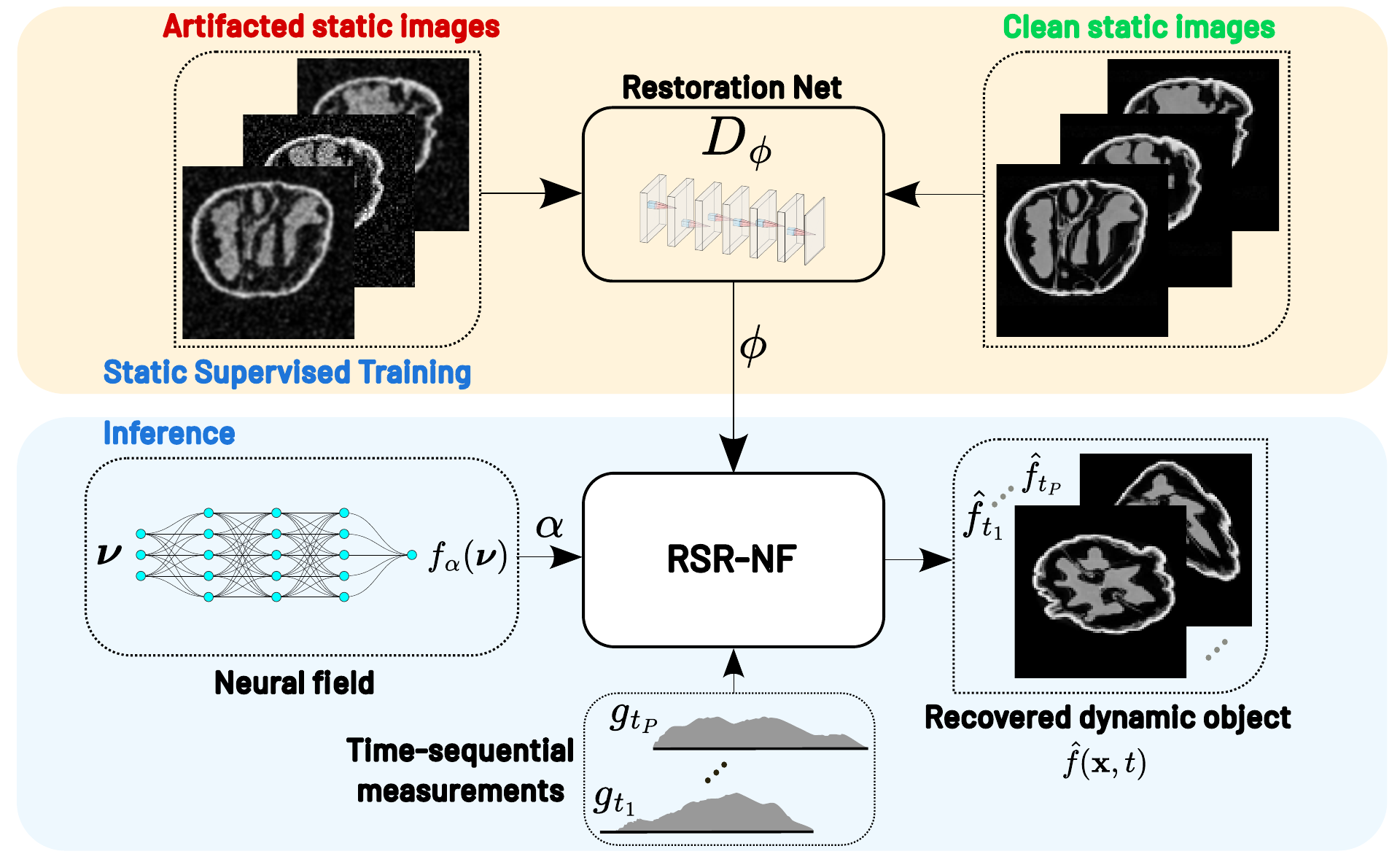}
    \setlength{\abovecaptionskip}{0pt}
    \setlength{\belowcaptionskip}{-8pt}
    \caption{
    \small 
    The RSR-NF framework. The deep {restoration network} $D_\phi$ is trained on slices of \textit{static} objects similar to the object of interest, and the learned spatial prior is used at inference time.
    }
    \label{fig:framework}
\end{figure}

\subsection{Neural field representation for the object $f$}
To represent the complete 2D dynamic object $f$, we use a {standard} neural field $N_{\alpha}: \mathbb{R}^{3} \rightarrow \mathbb{R}$ where the temporal coordinate is also added to the input as $\boldsymbol{\nu} = (\boldsymbol{x}, t)$
with all coordinates normalized such that $\boldsymbol{\nu} \in [0,1]^3$, and $\alpha$ represents the learnable parameters. The NF has a MLP architecture using fixed Fourier positional encodings $\gamma(\boldsymbol{\nu}) \in \mathbb{R}^{L \times 6}$ with $L$ linearly increasing frequencies {\cite{sun2021coil} to mitigate spectral bias \cite{rahaman2019spectral}}, where the $l$-th row is defined as $\gamma(\boldsymbol{\nu})_l = ( \sin( \pi l /2 \boldsymbol{\nu} ), \cos( \pi l /2 \boldsymbol{\nu} ) )$ where $l \in \{ 1, \dots, L \}$, and $\sin$ and $\cos$ are applied to coordinate vectors element-wise. 
{The positional encodings are followed by a} MLP network $M_{\alpha}: \mathbb{R}^{L \times 6} \rightarrow \mathbb{R}$ mapping the encoded coordinates to the estimated output densities.
Thus, the overall NF model computes the density values 
$\mu_{\boldsymbol{\nu}} \in \mathbb{R}$ at a given spatio-temporal coordinate $\boldsymbol{\nu} \in \mathbb{R}^{3}$ as 
\begin{align}
f^{\alpha}(\boldsymbol{\nu}) = M_\alpha (\gamma(\boldsymbol{\nu})) = \mu_{\boldsymbol{\nu}}. \nonumber
\end{align}

\subsection{Variational Formulation and Optimization}
The variational objective to be optimized is
\begin{align}
\label{eq:var_obj}
    \min_{\alpha, \bar{f}} &\sum_t \| g_{\theta(t), t} - R_{\theta(t)}\tilde{f}^{\alpha}_{t} \|_2^2 + \lambda \rho(\bar{f}_t) + \xi \rho_\tau(\tilde{f}^\alpha) \nonumber \\ 
    & \text{s.t.} \,\, \tilde{f}^\alpha = \bar{f} 
\end{align}
where $g_{\theta(t),t} \in \mathbb{R}^{J}$ is the {measured} time-sequential projection at time $t$ and view angle $\theta(t)$ corrupted with AWGN, $\tilde{f}^\alpha \in \mathbb{R}^{J^2 \times P}$ is the neural field representation of the object {$f^\alpha$} {rendered}
on a {fixed}, {spatially-vectorized} $J^2 \times P$ grid {aligned with the space-time coordinates}, and $\tilde{f}^\alpha_t \in \mathbb{R}^{J^2}$, $t=0, \ldots P-1$ is its $t$-th time instant {(``frame")}, and
$R_{\theta(t)}: \mathbb{R}^{J^2} \rightarrow \mathbb{R}^J$ is the {discretized} Radon transform operator at view angle $\theta(t)${. The temporal regularizer $\rho_\tau$  is a finite difference approximation to the energy in} the second order derivative with respect to time $\| \partial^2 f/\partial t^2 \|_F^2$, 
\begin{align}
\rho_\tau(\tilde{f}^\alpha) = \sum_{t=2}^{P-1} \| \tilde{f}^\alpha_{t-1} -2\tilde{f}^\alpha_t + \tilde{f}^\alpha_{t+1} \|_2^2. \nonumber 
\end{align}
The spatial regularizer $\rho$ with weight $\lambda > 0$ is the RED term
\begin{align}
\label{eq:red_reg}
    \rho(\bar{f}_t) = \bar{f}_t^T(\bar{f}_t - D_\phi(\bar{f}_t))
\end{align}
where $D_\phi: \mathbb{R}^{J^2} \rightarrow \mathbb{R}^{J^2}$ is a {restoration operator} that {takes} a single time {frame $f_t$ of the object} {as input} {and produces its restored version as output}.
{RED \cite{romano2017little} was originally inspired by the PnP method \cite{venkatakrishnan2013plug, kamilov2023plug}, which uses a denoiser to replace a proximal mapping and proposed {the explicit regularizer \eqref{eq:red_reg} with a denoiser $D$.} 
RED has been widely used for static reconstruction tasks \cite{metzler2018prdeep, sun2019block, mataev2019deepred}, and was recently applied to dynamic imaging \cite{iskender2023red, iskender2023redtci}. The original theoretical analysis \cite{romano2017little} has assumptions that are not simultaneously satisfied for many denoisers but the score-matching framework in \cite{reehorst2018regularization} explains good performance in such cases. 

Similar to recent works \cite{liu2020rare, hu2023restoration, hu2024stochastic, vo2024neural}, we replace the denoiser with a {learned deep NN} restoration operator $D_{\phi}$.}
{(See Section~\ref{sec:red_restorator} for further details of $D_{\phi}$.)}
{Note that in \eqref{eq:var_obj} and \eqref{eq:red_reg}, the spatial regularizer $\rho$ is applied to auxiliary variable $\bar{f}$, rather than directly to the NF output $\tilde{f}^\alpha$.} 
{However,} using the NF representation for the RED updates in solving the optimization problem \eqref{eq:var_obj}
{would require} costly {backpropagation computations} through the deep {restoration NN {for updating the NF parameters.}}
{To avoid this,} we perform the variable splitting $\tilde{f}^{\alpha} = \bar{f}$, {and in \eqref{eq:var_obj} and \eqref{eq:red_reg}, the spatial regularizer $\rho$ is applied to auxiliary variable $\bar{f}$, rather than directly to the NF output $\tilde{f}^\alpha$.}
The split variable $\bar{f} \in \mathbb{R}^{J^2 \times P}$ is constrained to equal the {rendered} NF representation $\tilde{f}^{\alpha}$.

To solve the resulting problem, we use an ADMM framework similar to \cite{Iskender_2023_ICCV, iskender2023red}. This time the variable split is performed on the NF representation instead of the low-rank bilinear PSM. The resulting augmented Lagrangian in the scaled form \cite{gabay1976dual, eckstein1992douglas} is
\begin{align}
\label{eq:aug_lag_nf}
\mathcal{L}_\beta[\tilde{f}^{\alpha}, \bar{f}; \gamma] = &\sum_t \Big( \Big\| R_{\theta_t} \tilde{f}^{\alpha} e_t - g_t \Big\|_2^2 + \lambda \rho(\bar{f} e_t) \Big) \nonumber \\
+ \xi \rho_\tau(\tilde{f}^\alpha)& - %
{\frac{\beta}{2}\| \gamma \|_F^2} + \frac{\beta}{2}\| \tilde{f}^\alpha - \bar{f} +{\gamma}\|_F^2,
\end{align}
where $\gamma \in \mathbb{R}^{J^2 \times P}$ is the dual variable and $\beta>0$ is the augmented Lagrangian weight. 

The proposed RSR-NF algorithm minimizes \eqref{eq:aug_lag_nf} with respect to the primal variables $\tilde{f}^\alpha$ and $\bar{f}$ and performs a dual ascent step for $\gamma$. The method is described in Algorithm \ref{alg:admm_nf_red}.
\begin{algorithm}[b]
\caption{RSR-NF}\label{alg:admm_nf_red}
\textbf{input:} $\alpha^{(0)}$, $\bar{f}^{(0)} = \tilde{f}_\alpha^{(0)}$, $\gamma^{(0)}$, $\beta>0$, $\lambda>0$, $\xi>0$\\
\begin{algorithmic}[1]
\FOR{$i\in\{1,\ldots,I\}$}
    \STATE $\alpha^{(i)} =
    \arg\min_{\alpha} \{ \sum_t \| g(\cdot, \theta(t), t) - R_{\theta(t)}\tilde{f}^{\alpha}_{t} \|_2^2 $ \par $ + \frac{\beta}{2}\| \tilde{f}^{\alpha}_{t} + %
    {\gamma_t^{(i-1)} - \bar{f}_t^{(i-1)}} \|_F^2 + \xi\rho_\tau(\tilde{f}^\alpha) \}$ \label{line:g_nf_step_nf_fidelity_alg1}
    \smallskip

    \STATE $\forall t:\,\, \bar{f}_t^{(i)} = \arg\min_{\bar{f}_t} \{ \lambda \rho(\bar{f}_t) $ \par $+ \frac{\beta}{2}\| (\tilde{f}^{\alpha (i)} + \gamma^{(i-1)})e_t - \bar{f}_t  \|_2^2 \}$
    \label{line:g_step_pnf_fidelity_alg1}
    \smallskip
    
    \STATE $\gamma^{(i)} = \gamma^{(i-1)} + {\tilde{f}^{\alpha(i)}} - \bar{f}^{(i)}$ \label{line:dual_step_nf_fidelity_alg1}
\ENDFOR
\end{algorithmic}
\end{algorithm}
{Step 2 of the algorithm is solved iteratively using gradient descent for a fixed number of updates per each outer iteration. The number of outer iterations $I$ is selected sufficiently high for convergence or marginal improvements with additional iterations.}

{Step 3} can be replaced by a fixed-point update with early stopping \cite{iskender2023red, Iskender_2023_ICCV, iskender2023dynamic}. Using the gradient rule 
\begin{align}
\nabla \rho(\bar{f}_t) = \bar{f}_t - D_\phi(\bar{f}_t), \nonumber
\end{align}
this leads to the efficient single use of the {restoration NN} $D_\phi$ at each outer iteration {in Algorithm~\ref{alg:admm_psm_red}}.
\begin{algorithm}[htb!]
\caption{RSR-NF with efficient $f$-step}
\label{alg:admm_psm_red}
\small{\textbf{Notes:} Inputs, and Lines 1-2 and 4-5 are the same as Algorithm \ref{alg:admm_nf_red}. 
The $\bar{f}$-step is applied $\smash{\forall t}$.}
\begin{algorithmic}[1]
\setcounter{ALC@line}{2}
    \STATE \normalsize{$\forall t:\,\, \bar{f}_t^{(i)} =  \frac{\lambda}{\lambda + \beta} D_\phi(\bar{f}_t^{(i-1)}) +
    {\frac{\beta}{\lambda + \beta} \Big( \tilde{f}^{\alpha(i)}_{t} +  \gamma_t^{(i-1)} \Big)}$ \label{line:g_step_nf_fidelity_alg2}}
\end{algorithmic}
\end{algorithm}

\subsection{Learning static restoration priors}
\label{sec:red_restorator}
Various strategies have been proposed to utilize restoration priors in different settings. As an unsupervised alternative, RARE \cite{liu2020rare} trains an image prior using reconstructions obtained from undersampled measurements and replaces the {deep denoiser in the RED \cite{romano2017little} framework by an artifact removal NN.} {However,} RARE may struggle for severely undersampled problems such as dCT with time-sequential measurements without assumptions such as clustering temporally adjacent projections. {Instead, RSR-NF uses supervised pre-training on relevant ground-truth static data.}

{The replacement of the denoiser in a RED or PnP framework by a restoration operator is motivated by} recent work {\cite{hu2023restoration, hu2024stochastic}}
that has shown that pre-trained restoration models can function as effective priors for unseen inverse problems, performing on par or better than the traditional {or even diffusion model-based denoisers in RED and PnP.}

Similar to the {regularization denoisers} in \cite{sun2019block, iskender2023red, Iskender_2023_ICCV}, the {restoration {deep NN in RSR-NF}} $D_\phi$ has a DnCNN architecture \cite{zhang2017beyond}. {If available,} the {NN}  is pre-trained on a training set of 2D slices of similarly distributed \textit{static} objects. 
In the absence of such data, {if the object is dynamic for a period of time and static otherwise,} 
the training is performed using pre and post-motion static slices of the same object.
This strategy is agnostic to the specific dynamic behavior of the object of interest. The variational objective for the supervised pretraining of the {NN} is
\begin{align}
\label{eq:red_denoiser_train}
\min_{\phi} \sum_i \| f_i - D_\phi (\hat{f}_i) \|_F^2 \,\, s.t. \,\, \hat{f}_i = H_i f_i + \eta_i, \,\, \forall i,
\end{align}
where to allow the {NN} is to correct various levels of artifacts and hence to diversify the ensemble of restoration scenarios, {similar to \cite{delbracio2023inversion, hu2024stochastic}, we set $H_{i}$ to a convex combination of the identity map $I$ and the degradation operator $G_i$ as $H_i = \zeta_i G_i + (1-\zeta_i) I$ where $\zeta_i \sim U[0,1]$. 
We select $G_i$ as Gaussian blur operator with kernel size $k_i$, and to further diversify the ensemble of restoration scenarios, we randomize the kernel size $k_{i} \sim U[0, k_{\max}]$. The injected noise $n_i \sim \mathcal{N}(0, \sigma_i^2 I)$ has noise level $\sigma_i \sim U[0, \sigma_{\max}]$.}
\section{Experiments}
\subsection{Datasets}

\textbf{$\quad\,\,$Compressed polymer.}
The dataset of a porous polymer under compression was {generated at} 
the Los Alamos National Laboratory {by physics-based numerical simulation} as the temporal evolution of the coordinates of finite element nodal positions.
The material is a polymer with a cellular structure under quasi-static compression where the complete temporal evolution consists of 1001 time frames. 
{The high computational load and the inability to compare the accuracy against real-world data prevents the use of multiple such simulations for supervised learning.}
For a voxel-based representation, we first create a 4D $128 \times 128 \times 128 \times 1001$ spatio-temporal grid with the object at $t=0$ at the center, and map each node in all time frames to the corresponding voxel. The mapping is binary, such that the voxel density is 1 if there are one or more nodes that are mapped to it, and 0 otherwise.
Then, we conduct our experiments on the $128 \times 128 \times 1001$ fixed 48th axial slice of the 3D dynamic object in Figure \ref{fig:ground_truth_frames_polymer}. 
Due to memory constraints, we partition the {1001} frames into {7} non-overlapping intervals of length $128$ + one of $105$ frames, and conduct experiments on the interval shown in Figure \ref{fig:ground_truth_frames_polymer}. This intermediate difficulty interval ($\#2$) was chosen after ranking {the intervals} based on the {accuracies of their NF embeddings using} a fixed architecture and the same number of iterations.

\textbf{Walnut dataset.}
We utilize reconstructions of two distinct (static) walnuts from the publicly accessible 3D walnut CT dataset \cite{der2019cone}. To generate a dynamic test object, we synthetically warp the central axial slice of one walnut using a sinusoidal piecewise-affine time-varying warp \cite{pwise_affine}. This 2D image is divided into an $N \times N$ uniformly spaced rectangular grid, and the vertical displacement of $\Delta_{n,t} = -C(t) \sin(3\pi n/N), n \in \{0, \ldots, N - 1\}$ is applied on each row independently to create the temporal warp, where $C(t)$ is a linearly increasing function of $t$ with $C(0) = 0$. The static axial, coronal, and sagittal slices of the other walnut are employed to train the restoration NN.

{For both datasets, time-sequential projection data is simulated from the dynamic object, and used as the measurement data in experiments as described in Section \ref{sec:exp_settings}.}

\begin{table*}[hbtp!]
    \small
    \setlength{\tabcolsep}{0.45pt}
    \renewcommand{\arraystretch}{0.0}
    \centering
    \begin{tabular}{cc}
    \vspace{-0.1cm}
    {\begin{sideways} $\,\,\,\,$ Poly. \end{sideways}} & \includegraphics[width=0.97\linewidth, trim={0 0.3cm 0 0.5cm},clip]{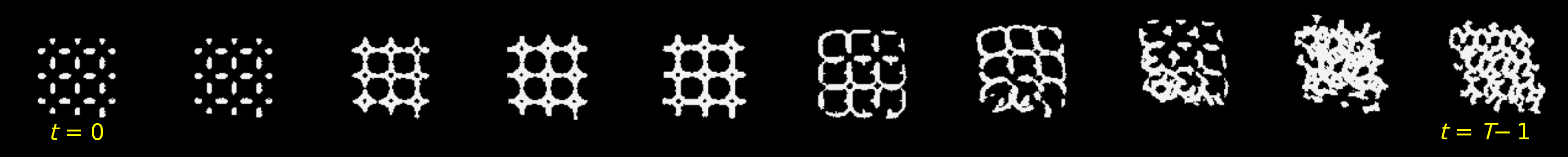} \\
    \vspace{-0.1cm}
    {\begin{sideways}$\,\,\,$Poly. int.
    \end{sideways}} & \includegraphics[width=0.97\linewidth, trim={0 0.3cm 0 0.5cm},clip]{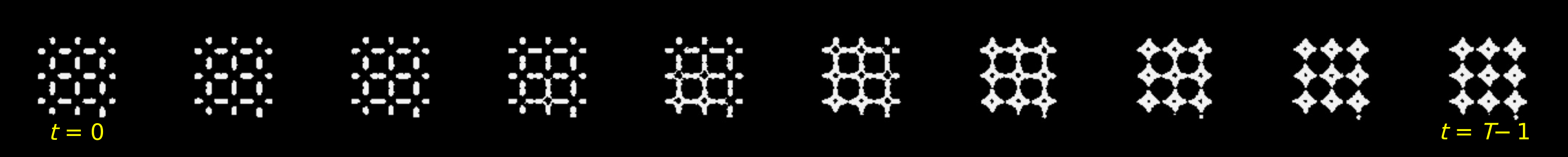} \\
    {\begin{sideways} $\,\,$ Walnut \end{sideways}} & \includegraphics[width=0.97\linewidth, trim={0.0cm 0.6cm 0.0cm 0.75cm}, clip]{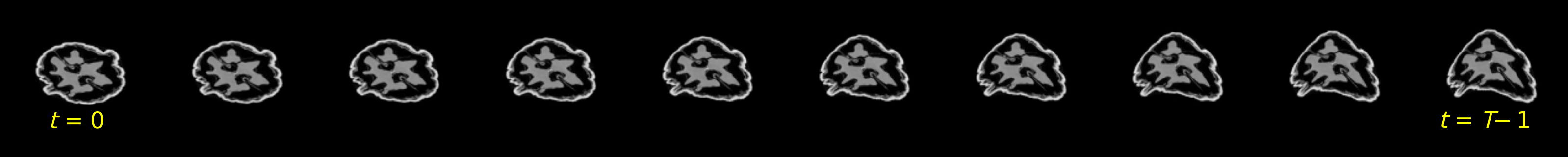} \\
    \end{tabular}
    \setlength{\abovecaptionskip}{4pt}
    \setlength{\belowcaptionskip}{-10pt}
    \captionof{figure}{\small
    {
    Ground-truth frames of different dynamic objects. Compressed polymer uniformly sampled in time
    {:
  full spatio-temporal object for $T$=1001 (top); the second non-overlapping subinterval with $T$=128 (center). Warped walnut for $T=128$ (bottom).
    }}}
    \label{fig:ground_truth_frames_polymer}
\end{table*}

\subsection{Comparison Benchmarks}

\textbf{$\quad\,\,$FBP:}
{To demonstrate the difficulty of the dynamic reconstruction problem, we directly reconstruct the object from its $P$ undersampled time-sequential projections using traditional filtered backprojection (FBP) with a Ram-Lak filter. Since FBP can only reconstruct a static instance, we use sliding windows of size $P/2$ with overlap $P/2$-1 for each {time frame}.}

\textbf{TD-DIP \cite{yoo2021time}:}
TD-DIP is an unsupervised DIP-based algorithm, proposed for dMRI. For dCT, we replace the forward operator with the 2D Radon transform $R_{\theta(t)}$. TD-DIP uses fixed latent representations to incorporate a motion-specific prior and subsequent two-stage deep neural network architecture to generate the estimated object at each time instant.

\textbf{RED-PSM \cite{Iskender_2023_ICCV,iskender2023redtci}:}
RED-PSM combines an object-domain low-rank PSM with RED \cite{romano2017little}, and the resulting optimization objective is minimized using a bilinear ADMM algorithm with convergence guarantees. 
To accelerate and improve convergence, the method can use a fast projection-domain PSM \cite{iskender2022dynamic, iskender2022dynamic_long} estimate for initialization.

\textbf{Temp-NF:}
As a key ablation study, we compare RSR-NF to {Temp-NF -- a version eliminating the spatial prior by setting $\lambda = 0$ in \eqref{eq:var_obj} but keeping the temporal regularization.}

\subsection{Experimental Settings}
\label{sec:exp_settings}
All experiments were conducted on a workstation with an Intel(R) Xeon(R) Gold 5320 CPU and NVIDIA RTX A6000 GPU. The optimization subproblem for the NF parameters in {Algorithm~\ref{alg:admm_psm_red}} was solved using SGD with the Adam \cite{kingma2014adam} optimizer. For each update, 
{mini-batches of $P/8$ projections were sampled at random (with replacement) from the entire set of $P$ projections.}

{\textbf{{Restoration Network} and Pre-training.}}
As in \cite{iskender2023red, Iskender_2023_ICCV, iskender2023dynamic}, each convolutional layer in {the restoration} CNN {$D_\phi$} is followed by a ReLU nonlinearity, except for the final single output channel layer. 
{One restoration NN} was pre-trained for each of the {two} objects. The upper limits for the noise level and blur kernel size for training {$D_\phi$} were set to $\sigma_{\max} = 5 \cdot 10^{-2}$ and $k_{\max} = 2$ pixels, respectively.
For the dynamic walnut object, $D_\phi$ is trained on the central 200 axial, 200 sagittal, and 200 coronal slices of another static walnut CT reconstruction downsampled to size $128\times128$. For the compressed polymer, a total of 422 axial, sagittal, and coronal slices of the pre- and post-compression objects were used for pre-training. Importantly, although we experiment on a temporal subinterval of the full dynamic object, we do not assume the availability of the beginning and end frames of the subinterval for training.

{\textbf{Tomographic Acquisition Scheme.}}
{The acquisition of projection data in the parallel beam geometry is simulated by {applying a numerical forward projection operator} to the dynamic object data at different time instants, to produce $P$ projections. To simulate a realistic noise level, the projections are corrupted  by additive white Gaussian noise  of intensity chosen such that a conventional Ram-Lak FBP reconstruction of the static version of the object from $P=512$ such projections will have a PSNR of 46 dB.}

All methods here employ the bit-reversed angular sampling scheme  over the range $[0, \pi]$ for the projection angles $\{\theta_p\}_{p=0}^{P-1}$, as described in Section \ref{sec:dct_prob}. {Compared to alternative time-sequential acquisition schemes, this scheme has been shown} \cite{iskender2022dynamic, iskender2022dynamic_long} {to provide better conditioning of the forward {operator}.}
Although the bit-reversed scheme requires increased rotation speed, in certain applications such as MRI with radial acquisition, CT scanner with electronic beam deflection \cite{kulkarni_electron_2021}, or in scenarios where the acquisition time {of a single projection} dominates the rotation time {(e.g., because of low source photon flux or a dense material)}, this does not constitute a problem. 
In addition, in physically constrained settings, the number of distinct views can be reduced and these views can be repeated periodically, {facilitating implementation by} multiple source-detector pairs, or by carbon nanotube sources \cite{spronk_evaluation_2021, xu_volumetric_2023}. {This scenario is studied in one of the experiments reported in Section~\ref{subsec:results}.}

{\textbf{NF architecture and initialization.}}
Unless stated otherwise, the NF used in the experiments consists of {$L=10$} Fourier positional encodings {for each of the input coordinates} with linearly increasing frequencies in the interval $[{\pi}/{2}, L{\pi}/{2}]$ \cite{sun2021coil} {followed by a} MLP architecture with 7 hidden layers of width 64 and ReLU nonlinearities with randomly initialized parameters $\alpha$. 
{The number of hidden layers were chosen as $h=7$ such that it provides sufficient expressivity for our experiments. Indeed, for $L$ sinusoidal encodings the maximum encoding frequency is $L\pi/2$, and for resolution $J=128$ of the sampling grid the expressivity analysis in \cite{yuce2022structured} predicts at least $\log_2(\frac{J\pi}{L\pi/2}) < 5$ hidden layers to achieve the maximum frequency on the grid without aliasing. Hence, this is satisfied with some margin by the choice of $h=7$.} 

{\textbf{Evaluation Metrics.}}
To evaluate the performance of the compared methods, we use the peak signal-to-noise ratio (PSNR) in dB, the structural similarity index (SSIM) \cite{wang2004image}, the mean absolute error (MAE), and the high-frequency error norm \cite{ravishankar2010mr} as $\text{HFEN}(f, f_r)~\triangleq~\| \text{LoG}(f)-\text{LoG}(f_r) \|_2$ where LoG is the rotationally symmetric Laplacian of Gaussian filter with a standard deviation of 1.5 pixels.

\subsection{Results}
\label{subsec:results}

{\textbf{$\quad\,\,$Embedding experiments.}}
We compare the representation accuracies of the proposed NF model with the partition of unity NF (POUnet) NN architecture \cite{lozenski2022memory} and with low-rank PSM for different {numbers of} degrees of freedom {(parameters in the NF model). We perform this comparison on full data (not projection measurements)} 
{of the compressed polymer data subinterval.}
For the PSM, we show the rank-$K$ {approximation} accuracies obtained by truncating the SVD of the object $f$ to {the first $K$ dominant components.} 
{For the {two} NF representations, we evaluate embeddings with different numbers of MLP layers and} channels per layer. 
{The NF parameters are optimized by} the Adam optimizer with $125$K iterations and learning rate of $5\cdot 10^{-3}$.

Qualitative comparison of representations at different time instants using the low-rank PSM and the proposed NF architecture is shown in Figure \ref{fig:embed_recon_est} for the polymer subinterval.
In the comparison, the NF uses 7 layers and 64 channels per layer, and the PSM has a rank of 3 which leads to a similar number of free parameters for both alternatives. As can be seen in the absolute error figures, the NF provides substantial accuracy improvements for both objects, whereas the PSM leads to over-smoothed representations.

We also provide PSNR (in dB) comparisons for various representations with different number of degrees of freedom in Figure~\ref{fig:embed_recon_est}. 
The results verify that for similar numbers of free parameters, NF-based representations with {various} widths and depths provide significant improvements over the PSM.
The NF architecture used in this work and the POUnet alternative {of} \cite{lozenski2022memory} provide similar representation accuracies for a similar number of free parameters.
However, {for polymer subinterval,} when the number of  degree of freedom is small, the proposed architecture provides improvement over the POUnet.

Another aspect in the comparison of the PSM vs. NF approaches, is that for each $K$, the accuracy for the {object SVD} truncated to rank $K$ constitutes an upper bound on the {$\ell_2$ sense} performance of the {rank-$k$} PSM. 
Conversely, {for the NF embeddings, changes in the optimization scheme may enable} further accuracy improvements.

\begin{table}[hbtp!]
\footnotesize
    \setlength{\tabcolsep}{0.45pt}
    \renewcommand{\arraystretch}{1.25}
    \centering
    \begin{tabular}{cccc}
    & {Embedding slices} & {Absolute errors} & 
    \\
    (a) & \includegraphics[align=c, width=0.258\textwidth, trim={0 0cm 0 0}, clip]{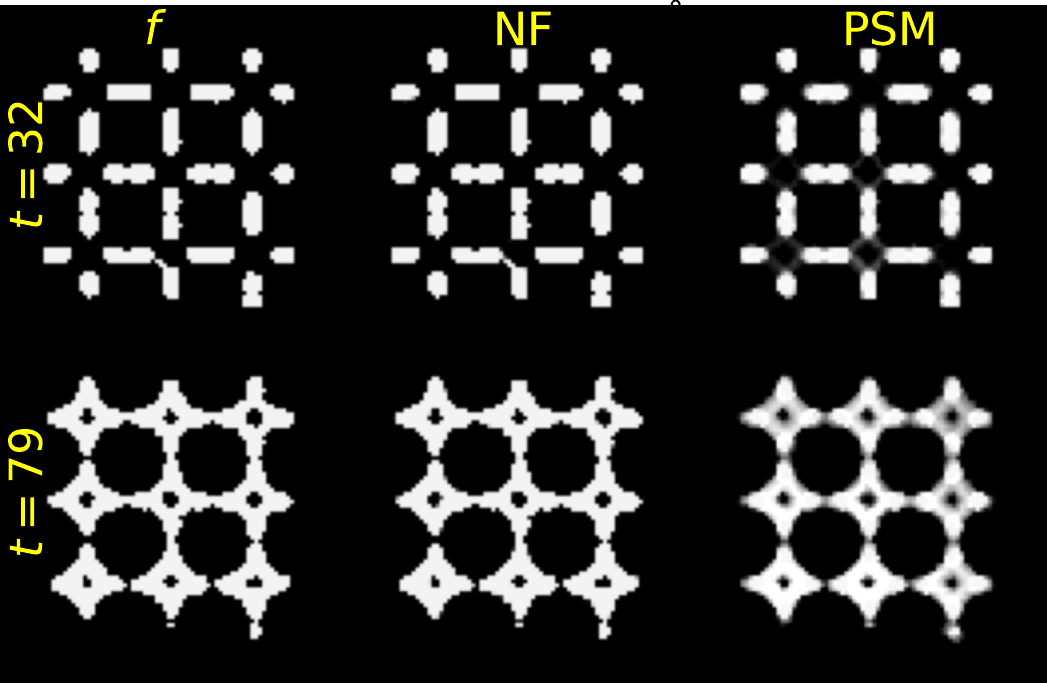} \hspace{-0.15cm} &
    \includegraphics[align=c, width=0.172\textwidth, trim={0.0 0cm 0 0}, clip]{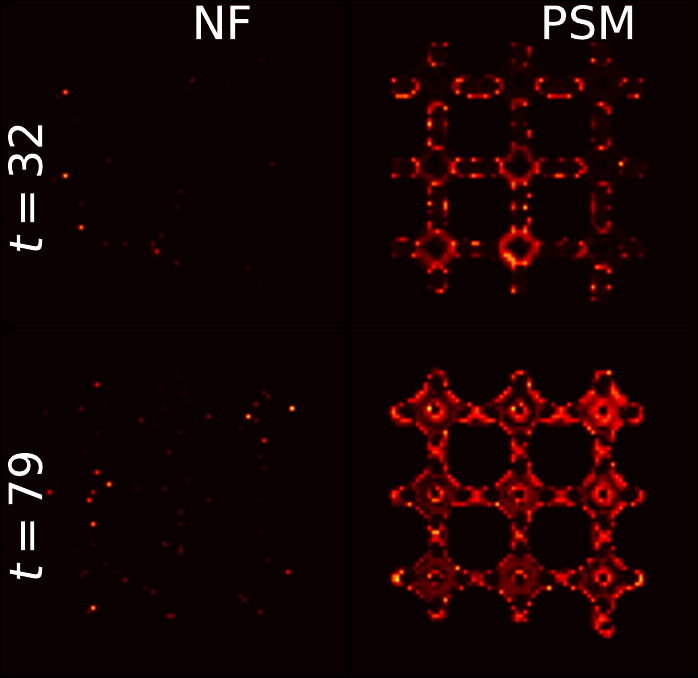} &
    \includegraphics[align=c, width=0.036\textwidth]{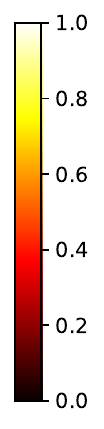} 
    \vspace{0.1cm}
    \end{tabular}
    \begin{tabular}{cc}
    (b)$\quad$ & \multicolumn{1}{c}{\includegraphics[align=c, trim={0 0 0 0.0cm}, clip, width=0.7\linewidth]{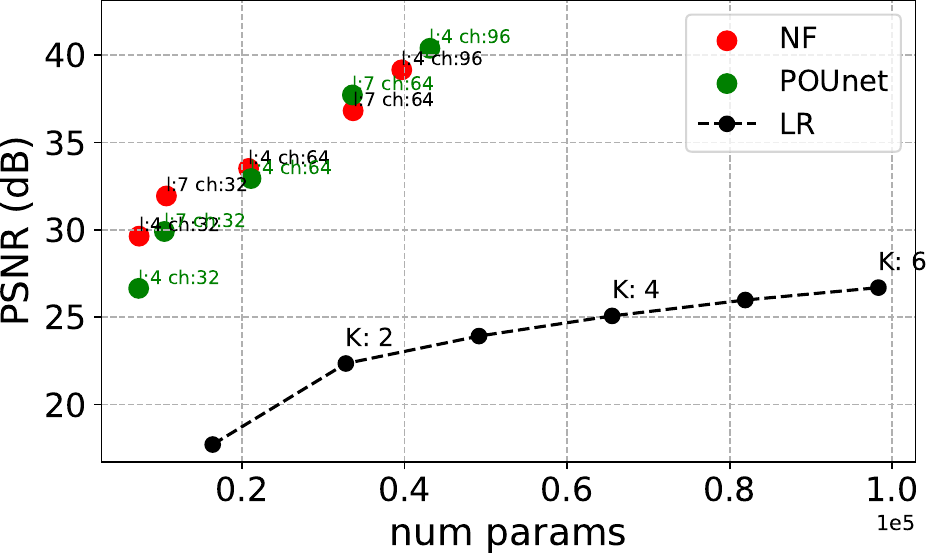}}
    \\
    \end{tabular}
    \setlength{\abovecaptionskip}{2pt}
    \setlength{\belowcaptionskip}{-8pt}
    \captionof{figure}{
    \small
    {(a) Representations {\& corresponding absolute errors for  $T$=128 polymer time frames} at two times for NF embedding with 7 layers and 64 channels per layer and for a rank-$K$=3 PSM{, which has roughly $1.5\times$ more parameters than the NF.}
    (b) Representation PSNR (in dB) for different NF architectures and rank-$K$ PSM.
    }
    }
    \label{fig:embed_recon_est}
\end{table}

{\textbf{dCT reconstruction accuracies for different $P$.}}
We compare RSR-NF with {four different benchmarks: (i)} FBP; (ii) TD-DIP \cite{yoo2021time}; (iii) RED-PSM \cite{Iskender_2023_ICCV, iskender2023red, iskender2023dynamic}; and (iv) Temp-NF. {This comparison is done} on the warped walnut and on the polymer subinterval, {using the time-sequential projection data}. 
{See Supplementary Material Table \ref{tab:avg_rec_acc_comp_params_rednf} for} the parameter configurations for {the various} methods.
{The accuracy metrics {of the various reconstructions} vs. number of time-sequential measurements $P$ are shown in Figure~\ref{fig:walnut_metrics_errbar_vs_P}.}
{For each $P$, the accuracies are calculated only based on the $P$ frames from which the projections are obtained.}
A more precise comparison of the plotted accuracies are provided in Table \ref{tab:avg_rec_acc_comp_walnut}.

For both objects and for various $P$, the RSR-NF reconstructions {with learned static spatial priors} have substantially improved metrics over {RED-PSM and TD-DIP, and also improve consistently over {the ablated version} with only temporal regularization, Temp-NF.}

{To visually assess the temporal resolution of the different methods, Figure~\ref{fig:recon_comp_walnut_slice} compares} reconstructed $x$-$t$ "slices" through the walnut and the compressed polymer, {taken at the locations indicated by a yellow line on the static $x$-$y$ frame at $t=0$.}
As highlighted {for the walnut by an arrow and for the polymer by the green box on the respective} ground-truth $x$-$t$ slices, RSR-NF has higher temporal resolution and outperforms the comparison benchmarks.

In Figure \ref{fig:recon_comp_walnut}, we provide a qualitative comparison of reconstructions for {$P=128$ at} three different time instants for the dynamic walnut. To facilitate comparison, we include only the three best-performing methods in the figure, RSR-NF, Temp-NF and RED-PSM. Compared to RED-PSM, RSR-NF suppresses the errors further. Specifically, {as shown in the zoomed-in region indicated by the yellow box}, the high-frequency features around the shell of the walnut are recovered with greater accuracy by RSR-NF, whereas the RED-PSM suffers from over-smoothing. On the other hand, {focusing on the zoomed-in region indicated by green boxes,}
RSR-NF provides better recovery of the {uniform} density regions of the walnut.

Likewise, Figure~\ref{fig:recon_comp_poly} {compares reconstructions of the polymer subinterval for $P=128$ at} three different time instants. Again, we include {the same three} methods, RSR-NF, Temp-NF and RED-PSM. As expected from the results in Figure~\ref{fig:walnut_metrics_errbar_vs_P}, RSR-NF improves over the other methods, particularly visible at $t=49$. To facilitate the comparison, we provide, in addition to the RSR-NF reconstructions, the absolute reconstruction error maps.

\begin{table}[hbtp!]
    \footnotesize
    \setlength{\tabcolsep}{0.45pt}
    \renewcommand{\arraystretch}{0.35}
    \centering
    \begin{tabular}{cc}
    (i) Walnut & (ii) Polymer \vspace{-0.025cm} \\
    \includegraphics[width=0.47\linewidth]{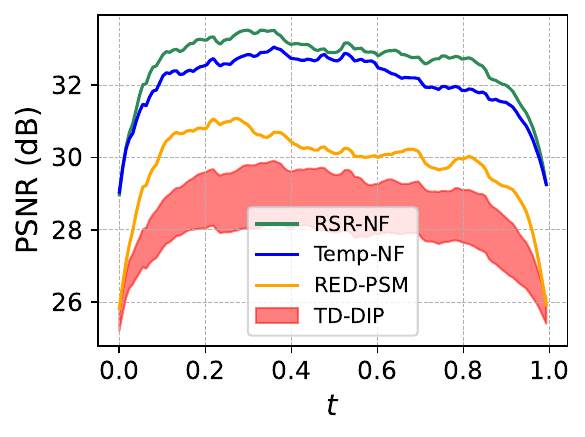} &
    \includegraphics[width=0.47\linewidth]{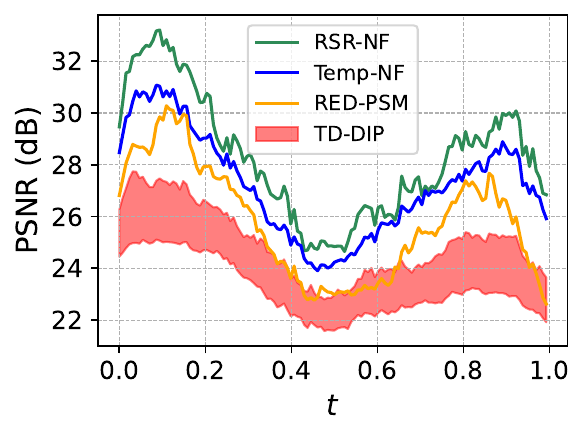} \vspace{-0.0cm} \\
    \end{tabular}
    \setlength{\abovecaptionskip}{-0pt}
    \setlength{\belowcaptionskip}{-4pt}
    \captionof{figure}{\small 
    Reconstruction PSNR vs. $t$ for (i) walnut and (ii) polymer. The shaded area for TD-DIP shows the interval between the best and the worst PSNR in three runs with random initialization.}
    \label{fig:metrics_vs_t_comp_and_small_period}
\end{table}

\begin{table}[hbtp!]
    \footnotesize
    \setlength{\tabcolsep}{0.45pt}
    \renewcommand{\arraystretch}{0.55}
    \centering
    \begin{tabular}{cc}
    { (i) Walnut} & { (ii) Polymer} \\
    \includegraphics[width=0.47\linewidth]{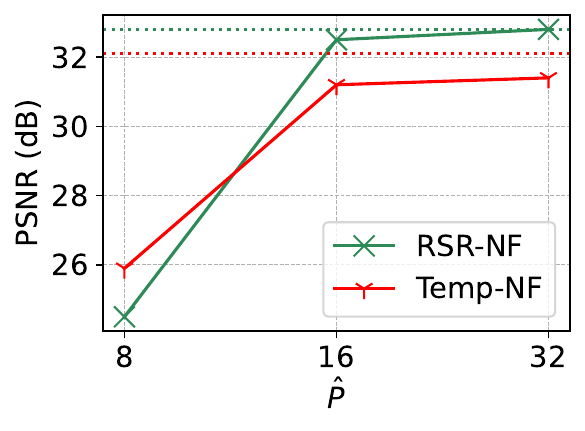} &
    \includegraphics[width=0.47\linewidth]{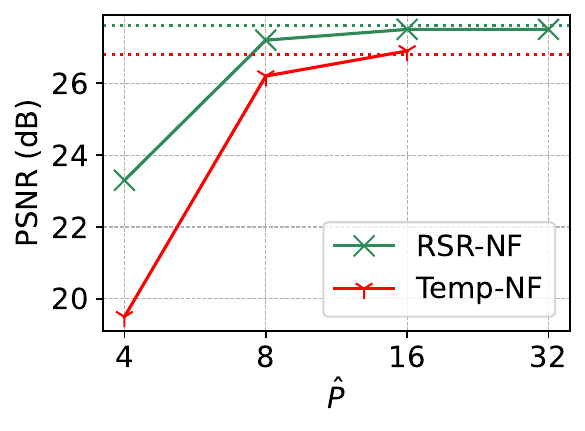} \\
    \end{tabular}
    \setlength{\abovecaptionskip}{-2pt}
    \setlength{\belowcaptionskip}{-4pt}
    \captionof{figure}{\small 
    Reconstruction PSNR vs. number of distinct view angles $\hat{P}$ with $P=128$ for (i) walnut, and (ii) polymer subinterval.}
    \label{fig:small_period}
\end{table}

    \begin{table}[hbtp!]
    \footnotesize
    \setlength{\tabcolsep}{4pt}
    \renewcommand{\arraystretch}{0.75}
    \centering
    \begin{tabular}{@{}cclcccc@{}}
    \toprule
    & \multicolumn{1}{c}{$P$} & \multicolumn{1}{l}{Method} 
    & \multicolumn{1}{c}{\makecell{PSNR (dB)}} & \multicolumn{1}{c}{SSIM} & \multicolumn{1}{c}{\makecell{MAE (1e-3)}} & \multicolumn{1}{c}{HFEN} \vspace{-0.05cm}\\
    \midrule
    \midrule
       & 32 
      & FBP & 15.1 & 0.300 & 8.4 & 2.13 \\
      & & TD-DIP (L) 
      & 22.5 & 0.882 & 2.1 & 0.87 \\
      & & RED-PSM 
      & {22.8} & {0.911} & {1.5} & {0.78} \\
      & & Temp-NF 
      & 23.7 & 0.937 & 1.4 & 0.71 \\
      & & \textbf{RSR-NF} 
      & \textbf{24.4} & \textbf{0.946} & \textbf{1.3} & \textbf{0.65} \\
     \midrule
     \multirow{5}{*}{\raisebox{-0.00cm}{\rotatebox[origin=c]{90}{\underline{Walnut}}}} & 64 
     & FBP & 18.3 & 0.387 & 5.4 & 2.04 \\
      & & TD-DIP (L) 
      & 25.6 & 0.916 & 1.5 & 0.74\\
      & & RED-PSM 
      & {26.4} & {0.958} & {0.9} & {0.57}\\
      & & Temp-NF
      & 28.0 & 0.974 & 0.8 & 0.49 \\
      & & \textbf{RSR-NF}
      & \textbf{29.4} & \textbf{0.982} & \textbf{0.7} & \textbf{0.40} \\
     \midrule
      & 128 
      & FBP & 20.3 & 0.503 & 3.8 & 2.22 \\
      & & TD-DIP (L) 
      & 29.0 & 0.953 & 1.0 & 0.50\\
      & & RED-PSM
      & {30.3} & {0.979} & {0.6} & {0.40}\\
      & & Temp-NF  
      & 32.1 & 0.988 & 0.5 & 0.33 \\
      & & \textbf{RSR-NF}  
      & \textbf{32.8} & \textbf{0.988} & \textbf{0.5} & \textbf{0.32} \\
      \midrule
      \midrule
      & 32 
      & FBP & 9.6 & 0.292 & 22.69 & 50.1 \\
      & & TD-DIP (L) 
      & 21.1 & 0.850 & 3.09 & 9.67 \\
      & & RED-PSM 
      & 20.2 & 0.865 & 2.42 & 10.92 \\
      & & Temp-NF 
      & 21.6 & 0.954 & 1.88 & 9.15 \\
      & & \textbf{RSR-NF} 
      & \textbf{22.3} & \textbf{0.967} & \textbf{1.41} & \textbf{7.83} \\
      \midrule
      \multirow{5}{*}{\raisebox{-0.00cm}{\rotatebox[origin=c]{90}{\underline{Polymer}}}} & 64 
      & FBP & 14.4 & 0.352 & 12.30 & 35.6 \\
      & & TD-DIP (L) 
      & 23.5 & 0.915 & 2.13 & 8.59 \\
      & & RED-PSM 
      & 22.7 & 0.917 & 1.51 & 9.74 \\
      & & Temp-NF
      & 24.1 & 0.977 & 1.20 & 8.22 \\
      & & \textbf{RSR-NF}
      & \textbf{24.8} & \textbf{0.983} & \textbf{0.84} & \textbf{7.33} \\
      \midrule
      & 128 
      & FBP & 17.1 & 0.466 & 7.64 & 31.49 \\
      & & TD-DIP (L) 
      & 24.3 & 0.849 & 2.41 & 10.58 \\
      & & RED-PSM
      & 25.3 & 0.940 & 0.98 & 9.68 \\
      & & Temp-NF 
      & 26.8 & 0.986 & 0.74 & 8.31 \\
      & & \textbf{RSR-NF} 
      & \textbf{27.6} & \textbf{0.988} & \textbf{0.55} & \textbf{7.35} \\
      \midrule
      \bottomrule
    \end{tabular}
    \setlength{\abovecaptionskip}{4pt}
    \setlength{\belowcaptionskip}{-6pt}
    \caption{
    \small 
    Reconstruction accuracies for different $P$ for dynamic walnut, and compressed polymer subinterval. For TD-DIP, the reported accuracies are for the best PSNR using a ``stopping oracle", averaged over three runs with random initial conditions.}
    \label{tab:avg_rec_acc_comp_walnut}
    \end{table}

    \begin{table*}[hbtp!]
    \small
    \setlength{\tabcolsep}{-1.4pt}
    \renewcommand{\arraystretch}{0.55}
    \centering
    \begin{tabular}{ccccc}
    {\raisebox{1.65cm}{\rotatebox[origin=c]{90}{\underline{Walnut}}}} & \includegraphics[width=0.25\linewidth]{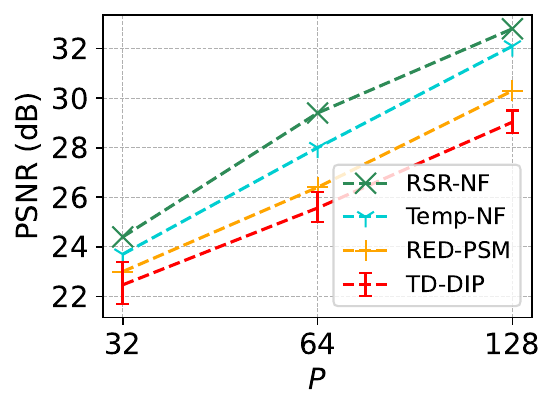} &
    \includegraphics[width=0.25\linewidth]{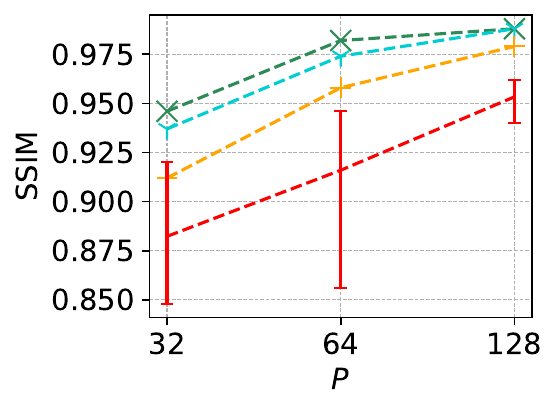} &
    \includegraphics[width=0.25\linewidth]{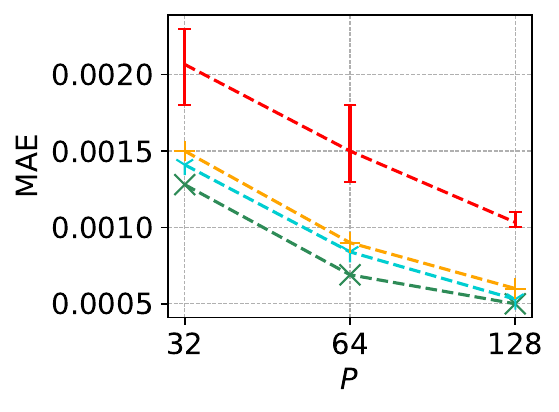} &
    \includegraphics[width=0.25\linewidth]{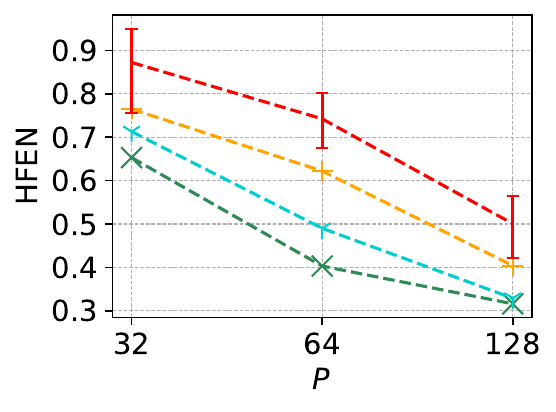} 
    \vspace{-0.675cm} 
    \\
    {\raisebox{1.65cm}{\rotatebox[origin=c]{90}{\underline{Poly. subint. 
    }}}} & \includegraphics[width=0.25\linewidth]{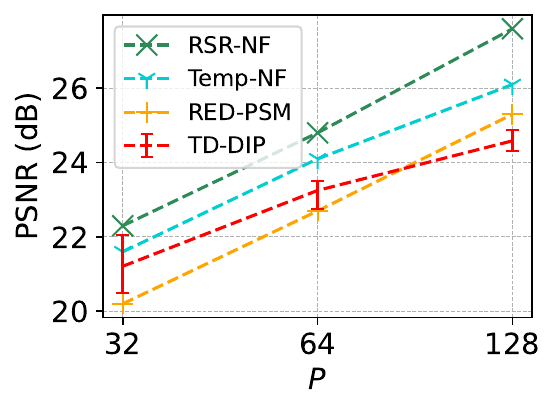} &
    \includegraphics[width=0.25\linewidth]{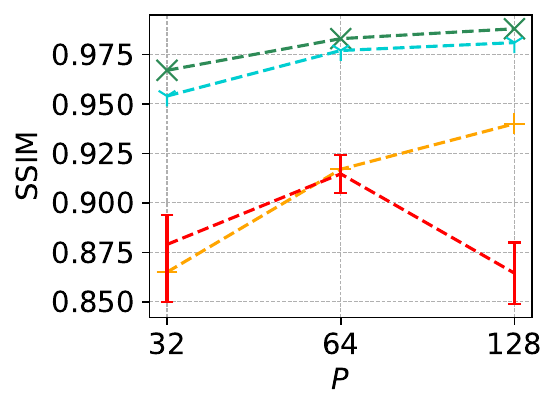} &
    \includegraphics[width=0.25\linewidth]{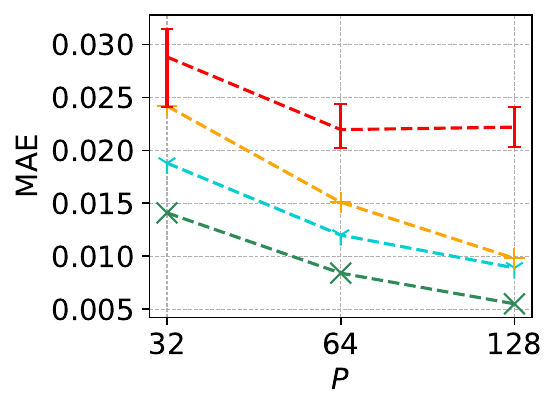} &
    \includegraphics[width=0.25\linewidth]{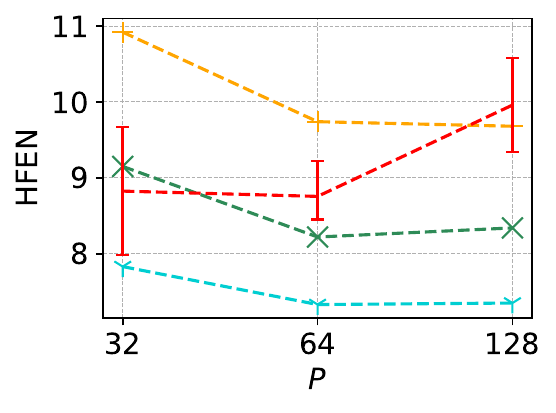}
    \end{tabular}
    \setlength{\abovecaptionskip}{-4pt}
    \setlength{\belowcaptionskip}{-2pt}
    \captionof{figure}{
    \small 
    {Reconstruction metrics for the time-varying walnut and compressed polymer vs. $P$ using different methods. For TD-DIP, the metrics reported are assuming a ``stopping oracle" that stops the iterations at the best PSNR reconstruction.}}
    \label{fig:walnut_metrics_errbar_vs_P}
    \end{table*}

\begin{table*}[hbtp!]
    \setlength{\tabcolsep}{0.45pt}
    \renewcommand{\arraystretch}{0.55}
    \centering
    \begin{tabular}{cc}
    \includegraphics[width=0.133\linewidth]{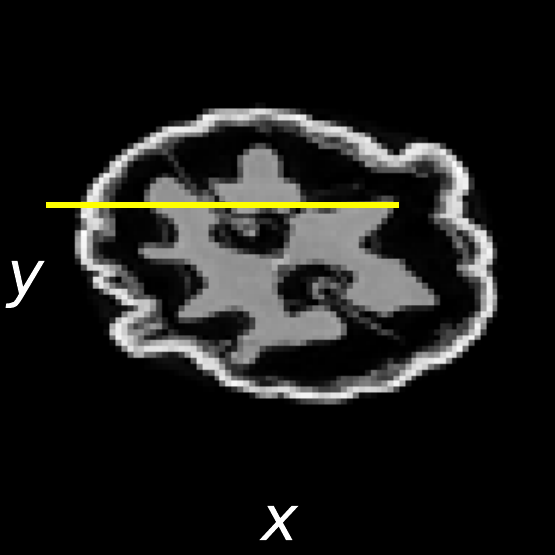} &
    \includegraphics[width=0.8545\linewidth]{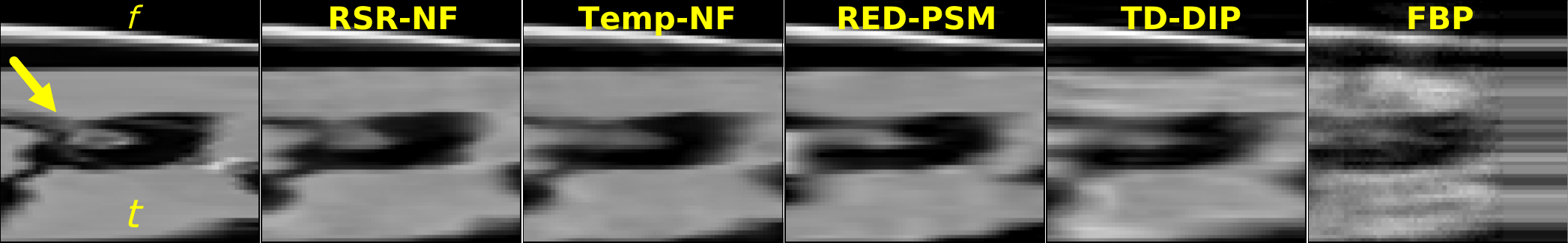}
    \end{tabular}
    \begin{tabular}{cc}
    \includegraphics[width=0.113\linewidth]{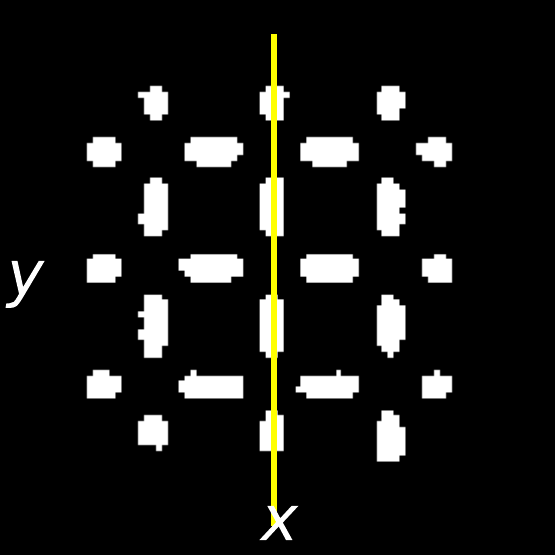} &
    \includegraphics[width=0.875\linewidth]{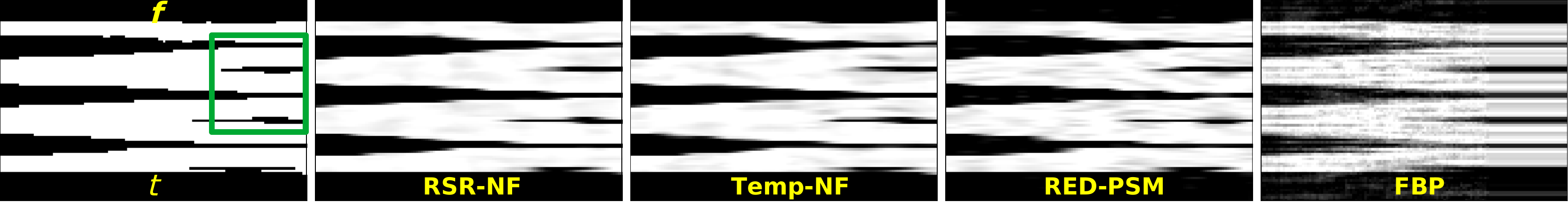}
    \end{tabular}
    \setlength{\abovecaptionskip}{2pt}
    \setlength{\belowcaptionskip}{-2pt}
    \captionof{figure}{
    \small 
    Comparison of reconstructed $x$-$t$ slices using different methods for dynamic walnut (top) and compressed polymer (bottom) for $P=128$. The cross-sections are shown on the static $t=0$ objects with a yellow line. The $x$-$y$ coordinates are indicated in white on the static objects and $t$ coordinate is shown on the slices of $f$ in yellow.
    }
    \label{fig:recon_comp_walnut_slice}
\end{table*}

\begin{table*}[hbtp!]
    \setlength{\tabcolsep}{0.45pt}
    \renewcommand{\arraystretch}{0.55}
    \centering
    \begin{tabular}{ccc}
    {\small Reconstructed slices} & {\small Zoomed-in (\textcolor{orange}{yellow} box)} & {\small Zoomed-in (\textcolor{ForestGreen}{green} box)}
    \\
    \includegraphics[width=0.2\linewidth]{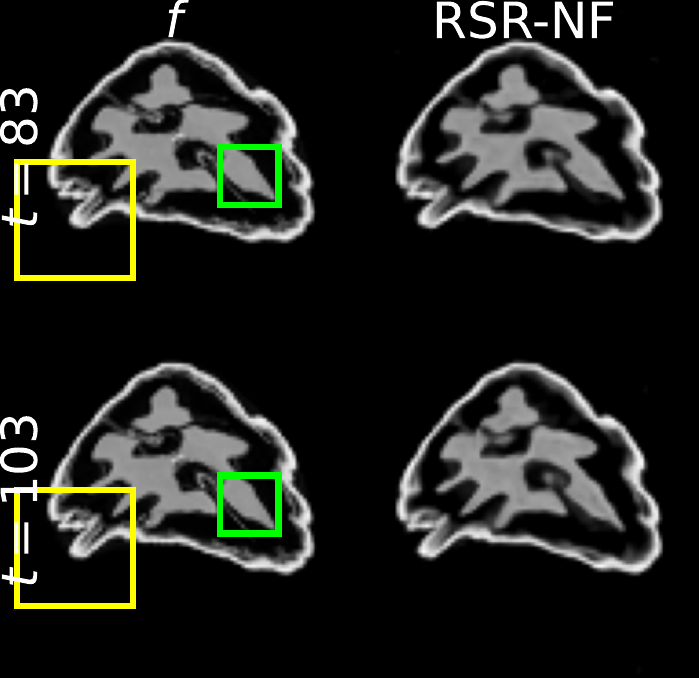} \hspace{-0.15cm} &
    \includegraphics[width=0.50\linewidth]{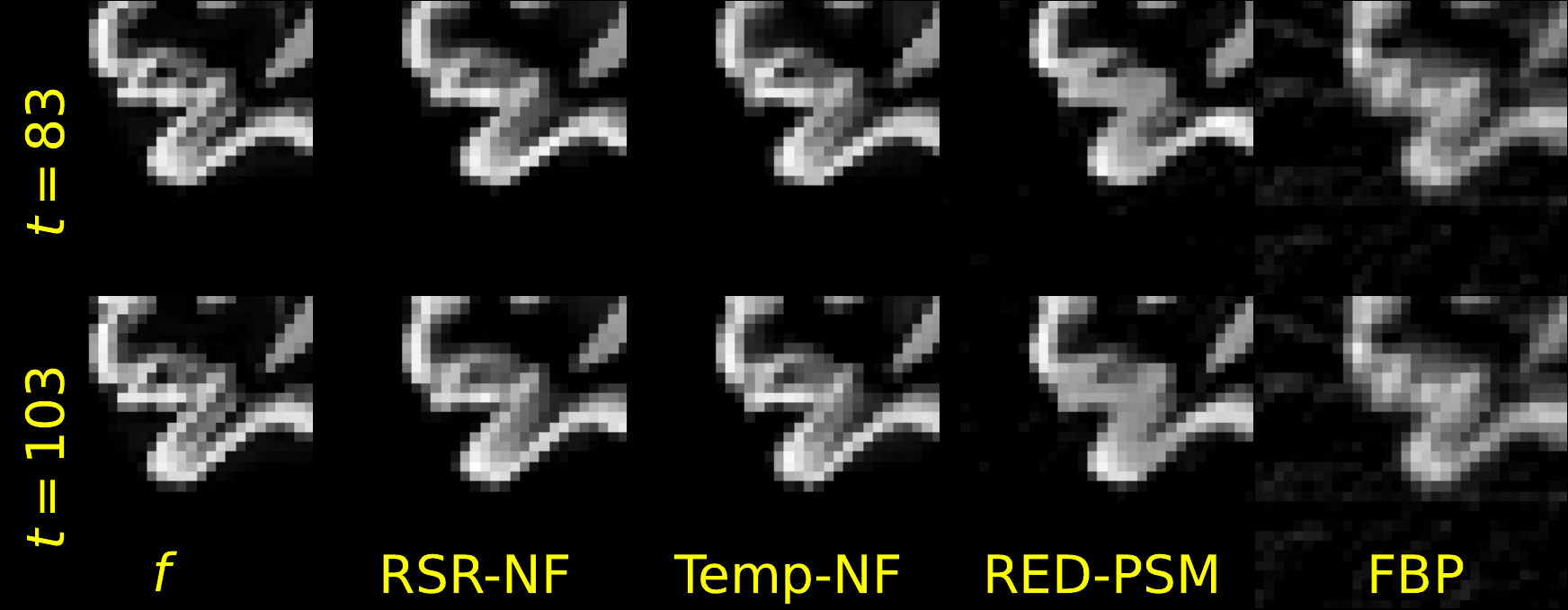} &
    \includegraphics[width=0.30\linewidth]{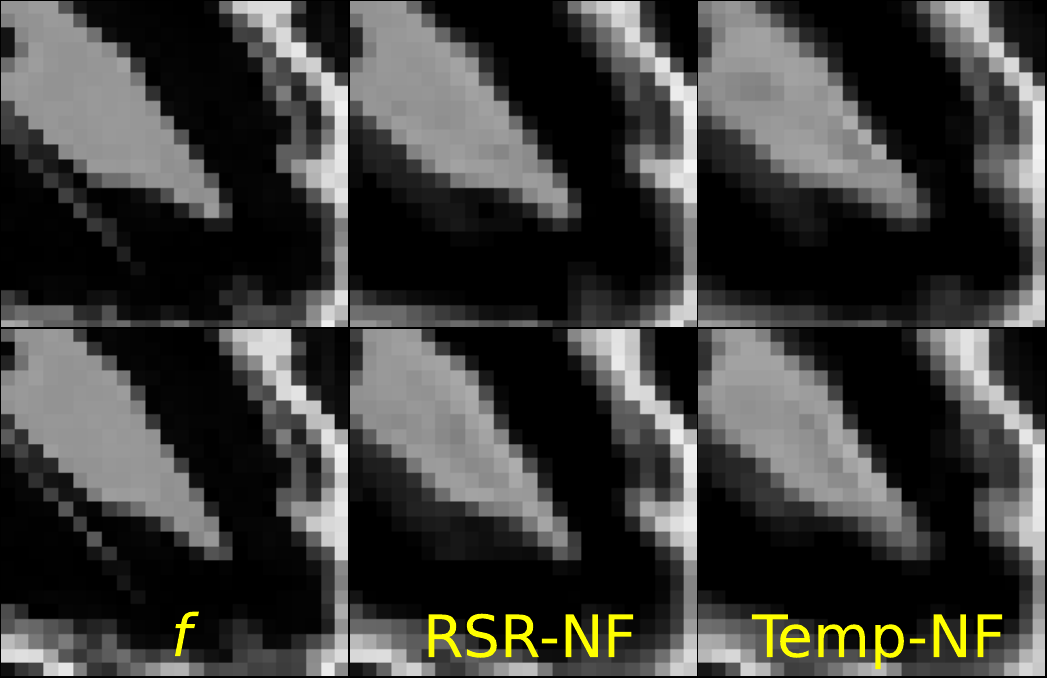}
    \end{tabular}
    \setlength{\abovecaptionskip}{2pt}
    \setlength{\belowcaptionskip}{-2pt}
    \captionof{figure}{
    \small 
    Comparison of reconstructed object frames for dynamic walnut at two time instants using different methods for $P=128$, and the corresponding normalized absolute reconstruction errors.
    }
    \label{fig:recon_comp_walnut}
\end{table*}

\begin{table*}[hbtp!]
    \setlength{\tabcolsep}{0.45pt}
    \renewcommand{\arraystretch}{0.55}
    \centering
    \begin{tabular}{ccc}
    {\small Reconstructed slices} & {\small Absolute reconstruction errors} & \vspace{0.00cm} \\
    \includegraphics[align=c, width=0.28\linewidth]{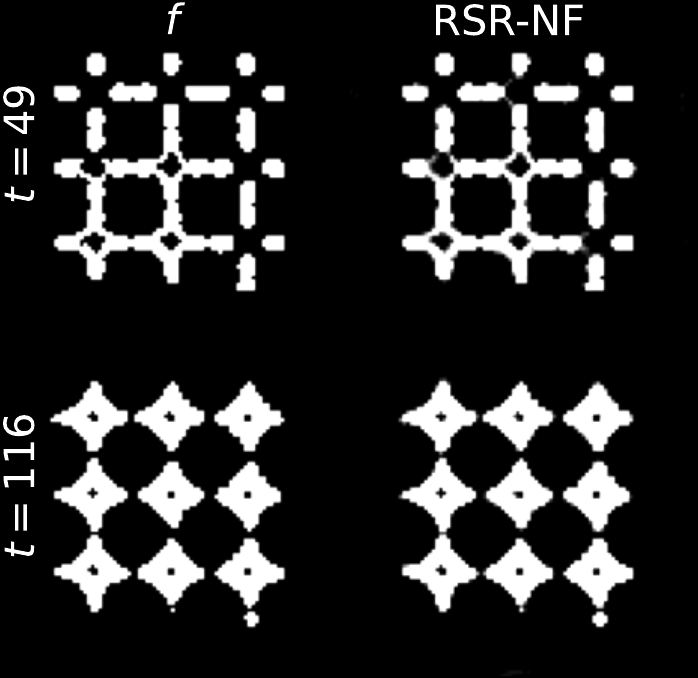} \hspace{-0.15cm} &
    \includegraphics[align=c, width=0.56\linewidth]{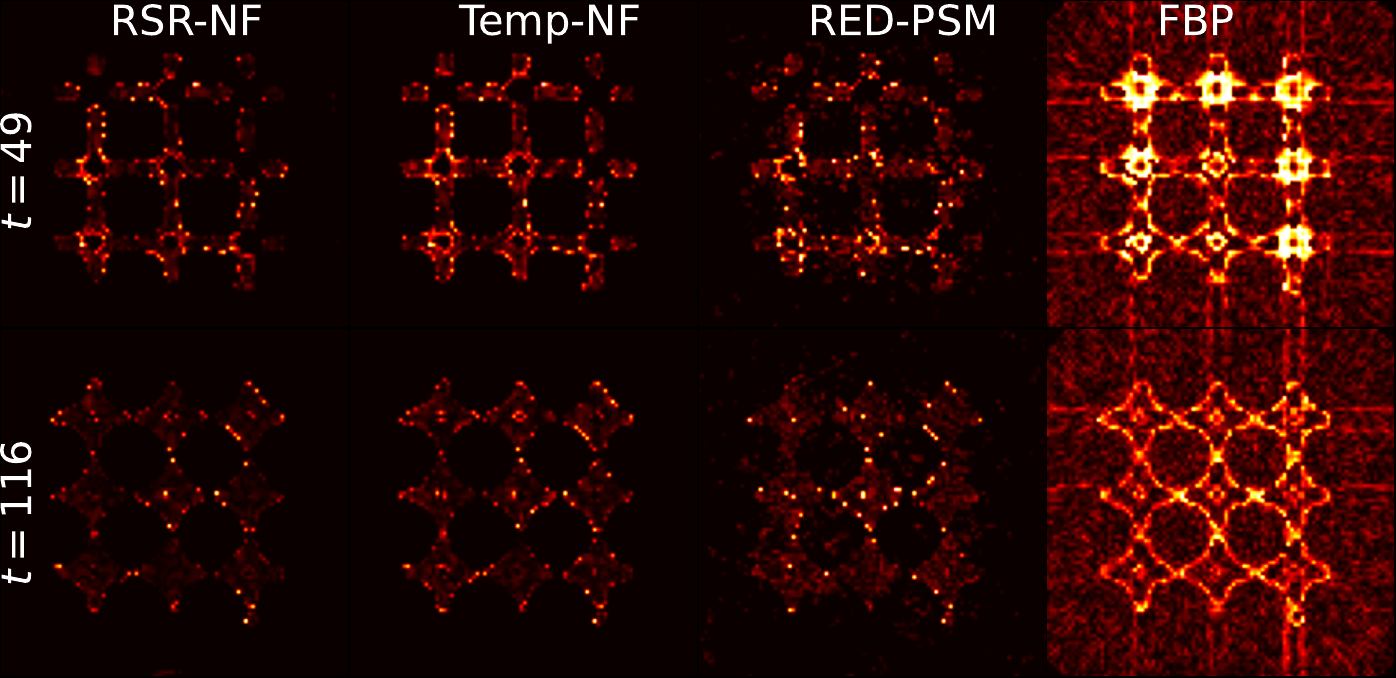} &
    \includegraphics[align=c, width=0.065\linewidth]{Figures/embedding/colorbar_recon.pdf}
    \end{tabular}
    \setlength{\abovecaptionskip}{4pt}
    \captionof{figure}{
    \small 
    Comparison of {ground-truth and RSR-NF} reconstructed object frames at two time instants for $P=128$, and the corresponding normalized absolute reconstruction errors {using different methods} for compressed polymer object subinterval.
    }
    \label{fig:recon_comp_poly}
\end{table*}

{\textbf{Smaller number of distinct view experiments.}}
{Since accessing a complete set of distinct view angles sequentially while using view angle schemes such as bit-reversed}
can be physically challenging, we also tested RSR-NF {and its ablated version without spatial priors, Temp-NF,}
in a setting {with only $\hat{P} < P$ distinct view angles that are sampled using the bit-reversed pattern, with the pattern repeated to produce a total of $P=128$ projections.} All other parameters were kept constant during the comparisons.

{The reconstruction PSNRs vs.} $\hat{P}$ for the two objects are shown in Figure~\ref{fig:small_period}.
The dashed lines indicate the performance of \textit{each method} in the ideal case of $\hat{P} = P = 128$. 
While RSR-NF is robust to the decrease in number of distinct view angles, Temp-NF {degrades more} compared to $\hat{P} = P $ for the walnut {for $\hat{P}>8$}. {At $\hat{P}=8$ the PSNR degrades considerably for both methods.} 
For the polymer, both methods show a similar trend compared to the $\hat{P} = P$ case, {but Temp-NF exhibits a sharper decline for $\hat{P}=4$.}

{These results show the potential of RSR-NF under physical constraints where the number of distinct view angles is reduced by eight or sixteen fold, with no performance loss.}

\section{Conclusions}
We proposed RSR-NF, {which is,} {to the best of our knowledge,} the first NF-based dynamic imaging method with pre-learned spatial priors. 
The proposed ADMM algorithm for optimization {of the variational objective is computationally efficient, by avoiding} costly {backpropagation} through the deep {regularizer NN} for {NF} updates. {Embedding} experiments show significantly better embedding {accuracy for a similar number of degrees of freedom by the NF than by a low-rank representation. Reconstruction experiments show improved  accuracies for RSR-NF compared to three benchmarks: a recent PSM-based alternative, RED-PSM; a DIP-based method; and a NF-based method without the learned spatial priors and with only temporal regularization.}
The dCT study highlights however the necessity for temporal regularization when the problem is severely undersampled. 

Future work may include exploration of different initialization types, comparison with recent NF-based methods proposed for different inverse problems {by adapting them to} dCT, and application of RSR-NF to different imaging scenarios.

{\footnotesize
\bibliographystyle{myIEEEtran}
\bibliography{bib, egbib, dCT, dMRI}}

\clearpage
\clearpage
\section*{Supplementary Material}
\subsection{Experimental configurations}
\label{supp:exp_config_rednf}
The parameter selections for the experiments listed in Table \ref{tab:avg_rec_acc_comp_walnut} 
are provided in Table \ref{tab:avg_rec_acc_comp_params_rednf}.  
Also, the architectural information for the DnCNN restoration networks used throughout this work is in Table \ref{tab:denoiser_arch_rednf}. Finally, we provide parameter configurations for reconstructions in Figure \ref{fig:small_period} with different number of distinct view angles $\hat{P}$ with $P=128$ for both objects in Table \ref{tab:param_config_small_P}. 

    \begin{table}[hbtp!]
    \footnotesize
    \centering
    \begin{tabular}{@{}cl|ccc|ccc@{}}
    \toprule
     \multicolumn{2}{c}{} & \multicolumn{3}{c}{(a) Walnut} & \multicolumn{3}{c}{(b) Polymer} \\
     \midrule
    \multicolumn{1}{c}{$P$} & \multicolumn{1}{c}{Method} & \multicolumn{1}{c}{$\lambda$} & \multicolumn{1}{c}{${\xi}$} & \multicolumn{1}{c}{$\beta$} & \multicolumn{1}{c}{$\lambda$} & \multicolumn{1}{c}{${\xi}$} & \multicolumn{1}{c}{$\beta$}\\
    \midrule 
      32 & Temp-NF & - & 1e+2 & - & - & 1e+2 & - \\
      32 & RSR-NF & 1e+0 & 1e+2 & 1e+0 & 1e+0 & 1e+2 & 1e+0  \\
     \midrule 
      64 & Temp-NF & - & 1e+2 & - & - & 1e+1 & - \\
      64 & RSR-NF & 1e-1 & 1e+2 & 1e-1  & 1e+0 & 1e+1 & 1e+0 \\
     \midrule 
      128 & Temp-NF & - & 1e+2 & - & - & 1e+2 & - \\
      128 & RSR-NF & 1e-1 & 1e+2 & 1e-1 & 1e+0 & 1e+2 & 1e+0  \\
     \bottomrule 
    \end{tabular}
    \caption{The parameter selections for the reconstructions in Table \ref{tab:avg_rec_acc_comp_walnut}.}
    \label{tab:avg_rec_acc_comp_params_rednf}
    \end{table}
    
    \begin{table}[hbtp!]
    \footnotesize
    \centering
    \begin{tabular}{@{}lcc@{}}
    \toprule
    \multicolumn{1}{c}{Dataset} & 
    \multicolumn{1}{c}{$\#$ of layers} & 
    \multicolumn{1}{c}{$\#$ of channels}\\
    \midrule
       Walnut & 6 & 64 \\
       Polymer & 6 & 64 \\
     \bottomrule
    \end{tabular}
    \caption{Restoration network DnCNN configurations for different datasets.}
    \label{tab:denoiser_arch_rednf}
    \end{table}

    \begin{table}[hbtp!]
    \footnotesize
    \centering
    \begin{tabular}{@{}cl|ccc|ccc@{}}
    \toprule
     \multicolumn{2}{c}{} & \multicolumn{3}{c}{(a) Walnut} & \multicolumn{3}{c}{(b) Polymer} \\
     \midrule
    \multicolumn{1}{c}{$\hat{P}$} & \multicolumn{1}{c}{Method} & \multicolumn{1}{c}{$\lambda$} & \multicolumn{1}{c}{${\xi}$} & \multicolumn{1}{c}{$\beta$} & \multicolumn{1}{c}{$\lambda$} & \multicolumn{1}{c}{${\xi}$} & \multicolumn{1}{c}{$\beta$}\\
    \midrule 
      4 & Temp-NF &  &  &  & - & 1e+2 & - \\
      8 & Temp-NF & - & 1e+2 & - & - & 1e+2 & - \\
      16 & Temp-NF & - & 1e+2 & - & - & 1e+2 & - \\
      32 & Temp-NF & - & 1e+2 & - & - & 1e+2 & - \\
      \midrule 
      4 & RSR-NF &  &  &  & 1e+0 & 1e+2 & 1e+0  \\
      8 & RSR-NF & 1e+0 & 1e+2 & 1e+0 & 1e+0 & 1e+2 & 1e+0 \\
      16 & RSR-NF & 1e+0 & 1e+2 & 1e+0 & 1e+0 & 1e+2 & 1e+0  \\
      32 & RSR-NF & 1e+0 & 1e+2 & 1e+0 & 1e+0 & 1e+2 & 1e+0  \\
     \bottomrule 
    \end{tabular}
    \caption{The parameter selections for the reconstructions with different number of distinct view angles $\hat{P}$ in Figure \ref{fig:small_period} for $P=128$.}
    \label{tab:param_config_small_P}
    \end{table}

\end{document}